\documentclass[prd,aps,a4paper,nofootinbib,twocolumn,showpacs,showkeys]{revtex4}

\usepackage{graphicx} 
\usepackage{latexsym,amsmath,amsfonts,amssymb}

\newcommand{\beq}{\begin{equation}}
\newcommand{\eeq}{\end{equation}}
\begin{document}

\title{High-energy hyperbolic scattering by neutron stars and black holes}

\author{Donato \surname{Bini}$^{1}$}
\author{Andrea \surname{Geralico}$^{1}$}

\affiliation{
$^1$Istituto per le Applicazioni del Calcolo ``M. Picone,'' CNR, I-00185 Rome, Italy
}

\date{\today}

\begin{abstract}
We investigate the hyperbolic scattering of test particles, spinning test particles and particles with spin-induced quadrupolar structure by a Kerr black hole in the ultrarelativistic regime. We also study how the features of the scattering process modify if the source of the background gravitational field is endowed with a nonzero mass quadrupole moment as described by the (approximate) Hartle-Thorne solution. 
We compute the scattering angle either in closed analytical form, when possible, or as a power series of the (dimensionless) inverse impact parameter. It is a function of the parameters characterizing the source (intrinsic angular momentum and mass quadrupole moment) as well as the scattered body (spin and polarizability constant). Measuring the scattering angle thus provides useful information to determine the nature of the two components of the binary system undergoing high-energy scattering processes.
\end{abstract}


\maketitle

\section{Introduction}

The gravitational scattering of a two-body system consisting of two black holes or a black hole and a neutron star has recently received much attention \cite{Glampedakis:2002ya,Shibata:2008rq,Sperhake:2008ga,Sperhake:2009jz,Sperhake:2012me}, in connection with the possibility to detect the associated emission of gravitational wave (GW) signals by the advanced phase of currently operating GW
Earth-based interferometers as well as (more likely) by future, forthcoming satellite missions involving  space-based interferometers (see, e.g., Refs. \cite{ligo,virgo,Abbott:2016blz,lisa}).		

In spite of the fact that the occurrence of hyperbolic encounters, in general, is expected to be as probable as that of coalescing phenomena,  
the analytical treatment of either case is not equally developed in the literature. For example, in the case of coalescence several analytical and semi-analytical  methods are available, in addition to the full numerical relativity (NR) simulations: Post-Newtonian (PN), Post-Minkowskian (PM), effective-one-body (EOB), perturbation theory and gravitational self-force (GSF), etc. \cite{Buonanno:1998gg,Buonanno:2000ef,Damour:2000we,Damour:2001tu,Schafer:2009dq,Blanchet:2013haa,Barack:2009ux,Poisson:2011nh,Detweiler:2008ft,Barack:2011ed}. 
The same methods can be applied, in principle, also to scattering processes, but with additional complications. For instance, the spectrum of the emitted waves in the latter case spreads all over the infinite range of all possible frequencies, whereas in the case of coalescence it is peaked around a single frequency (that of the inspiralling, quasi-circular motion) up to the very end of the process, i.e., the merger phase. This fact mainly explains why there are no analytical results coming from perturbation theory yet.

In general relativity, the scattering problem is well established and fully solved when one body has mass much smaller than the other so that backreaction effects can also be neglected, i.e., in the test-field approximation. In this regime the problem reduces to study the hyperbolic-like motion of a pointlike massive particle in a given background spacetime, starting at radial infinity along a certain asymptotic direction, reaching a minimum approach distance to the gravitational source, and then coming back to radial infinity being deflected from the original direction of an angle which is the main observable of the process.
For instance, if the source of the gravitational field is a black hole, the scattering angle can be expressed in terms of Elliptic integrals and is a function of the energy and angular momentum of the particle (a compendium can be found in the textbook of Chandrasekhar \cite{Chandrasekhar:1985kt}). 
A renewed interest in this problem can be found in the recent literature, with studies involving PN and PM expansions of the scattering angle \cite{Damour:2014afa,Damour:2016gwp,Hopper:2017qus,Bini:2017xzy,Bini:2017wfr,Vines:2017hyw}.

We have studied in previous works how the features of the scattering process modify if the scattered particle is no more pointlike, but is endowed with an internal structure given by its spin \cite{Bini:2017ldh,Bini:2017pee}. 
The motion is no longer geodesic, but accelerated due to the spin-curvature coupling, according to the the Mathisson-Papapetrou-Dixon (MPD) model \cite{Mathisson:1937zz,Papapetrou:1951pa,tulc59,Dixon:1970zza,ehlers77}.
The aim of the present paper is to compute the scattering angle in the high-energy limit by adding more structure to both the scattered body and the source of the gravitational field.
To this end, we will consider extended bodies with spin-induced quadrupole moment, using available results for quadrupolar particle motion in both Schwarzschild and Kerr spacetimes \cite{Bini:2013nw,Bini:2013uwa,Bini:2015zya}.
We will provide corrections to the scattering angle with respect to the spinless case up to the quadratic order in spin, also depending on the \lq\lq shape'' of the body, covering both cases of \lq\lq black hole-like'' and \lq\lq neutron star-like'' objects.
Furthermore, we will investigate the companion situation of a structureless particle moving along a hyperbolic-like geodesic orbit in the spacetime of a (slowly) rotating (slightly) deformed source endowed with a mass quadrupole moment, as described by the (approximate) Hartle-Thorne solution \cite{Hartle:1968si}.
We will provide corrections to the scattering angle with respect to the Kerr case which are linear in the quadrupole parameter of the source. 

We follow the notation and conventions of Ref. \cite{Misner:1974qy}. The signature of the metric is $+2$ and Greek indices run from 0 to 3, whereas Latin ones from 1 to 3.

\section{Hyperbolic-like equatorial motion in a Kerr spacetime}

Let us consider the Kerr spacetime, whose metric written in standard Boyer-Lindquist coordinates $(t,r,\theta,\phi)$ is given by
\begin{eqnarray}
\label{kerrmet}
ds^2 &=& -\left(1-\frac{2Mr}{\Sigma}\right)dt^2 
-\frac{4aMr}{\Sigma}\sin^2\theta dtd\phi+ \frac{\Sigma}{\Delta}dr^2\nonumber\\
&&+\Sigma d\theta^2+\frac{A}{\Sigma}\sin^2 \theta d\phi^2\,,
\end{eqnarray}
with $\Delta=r^2-2Mr+a^2$, $\Sigma=r^2+a^2\cos^2\theta$ and $A=(r^2+a^2)^2-\Delta a^2\sin^2\theta$.
Here $a$ and $M$ denote the specific angular momentum and the total mass of the spacetime solution, so that the quantity $\hat a=a/M$ is dimensionless.
The inner and outer horizons are located at $r_\pm=M\pm\sqrt{M^2-a^2}$.

The motion in the equatorial plane is governed by the geodesic equations  (see, e.g., Ref. \cite{Bini:2016ovy})
\begin{eqnarray}
\label{eqsmotionequatorial}
\frac{\Delta}{M^2}\frac{dt}{d\tau} &=& 
\left(\hat E-\frac{M^2}{r^2}\hat a\hat x\right)\left(\frac{r^2}{M^2}+\hat a^2\right)+ \frac{\Delta}{r^2}\hat a \hat x\,, \nonumber\\
\left(\frac{dr}{d\tau}\right)^2  &=& \left(\hat E-\frac{M^2}{r^2}\hat a\hat x\right)^2-\frac{\Delta}{r^2}\left(1+\frac{M^2}{r^2}\hat x^2\right)\,, \nonumber\\
\frac{\Delta}{M}\frac{d\phi}{d\tau}&=& 
\left(1-\frac{2M}{r}\right)\hat x +\hat a\hat E 
\,,
\end{eqnarray} 
where $\hat E$ and $\hat L$ are the conserved energy and azimuthal angular momentum per unit mass of the particle, respectively, and $x=\hat L-a\hat E$ (so that $\hat x=x/M$ is dimensionless).

The radial and azimuthal equations can be conveniently written in the following factorized form in terms of the dimensionless inverse radial variable $u=M/r$ 
\begin{eqnarray}
M^2\left(\frac{du}{d\tau}\right)^2&=&2\hat x^2u^4(u-u_1)(u-u_2)(u-u_3)\,,\nonumber\\
M\frac{d\phi}{d\tau}&=&\frac{2\hat x}{\hat a^2}u^2\frac{u_4-u}{(u-u_+)(u-u_-)}\,,
\end{eqnarray}
which can be combined to yield
\begin{eqnarray} 
\label{dudphi_geo}
\left(\frac{du}{d\phi}\right)^2&=&\frac{\hat a^4}{2}\frac{(u-u_+)^2(u-u_-)^2}{(u_4-u)^2}\times \nonumber\\
&& (u-u_1)(u-u_2)(u-u_3)\,.
\end{eqnarray}
Here $u_1<u_2<u_3$ are the ordered roots of the equation
\beq
u^3-(\hat x^2+2\hat a\hat E\hat x+\hat a^2)\frac{u^2}{2\hat x^2}+\frac{u}{\hat x^2}+\frac{\hat E^2-1}{2\hat x^2}=0\,,
\eeq
while
\beq
u_\pm=\frac{M}{r_\pm}\,,\quad
u_4=\frac{\hat L}{2M\hat x}
=\frac12\left(1+\frac{\hat a\hat E}{\hat x}\right)\,.
\eeq
For hyperbolic orbits we have $u_1<0<u\leq u_2<u_3$, with $u_2$ corresponding to the closest approach distance \cite{Chandrasekhar:1985kt}.

We are interested in the ultrarelativistic limit ($\hat E\gg1$) {\it for fixed values of the impact parameter} $b=\hat L/\sqrt{\hat E^2-1}$ (so that $\hat b=b/M$ is dimensionless).
In this limit the orbital equation (\ref{dudphi_geo}) becomes
\begin{eqnarray} 
\label{eqphidiu}
\frac{d \phi}{d u}&=&\pm\frac{\sqrt{2}}{\hat a^2}\frac{u_4-u}{(u-u_+)(u-u_-)}\times \nonumber\\
&&
\frac1{\sqrt{(u-u_1)(u_2-u)(u_3-u)}}
+O(\hat E^{-2})\,,
\end{eqnarray}
where the $\pm$ sign should be chosen properly during the whole scattering process (with $u_1<0<u\leq u_2<u_3$), depending on the choice of initial conditions.
Here $u_1<u_2<u_3$ are the ordered roots of the equation
\beq
2(1-\hat a\alpha)^2u^3-(1-\hat a^2\alpha^2)u^2+\alpha^2=0\,,
\eeq
with 
\beq
\alpha=\frac{1}{\hat b}\,,\qquad
\hat b=\frac{\hat L}{M\sqrt{\hat E^2-1}}\,,
\eeq
while
\beq
u_\pm=\frac{M}{r_\pm}\,,\qquad
u_4=\frac{1}{2(1-\hat a\alpha)}\,.
\eeq
They are given by 
\begin{eqnarray}
\label{roots}
u_1&=& \bar u\left[1-2\cos\left(\frac{2\Theta}{3}\right) \right]\,,\nonumber\\
u_2&=& \bar u\left[1+2\cos\left(\frac{2\Theta+\pi}{3}\right) \right]\,,\nonumber\\
u_3&=& \bar u\left[1-2\cos\left(\frac{2(\Theta+\pi)}{3}\right) \right]\,,
\end{eqnarray}
where
\beq
\bar u=\frac16\frac{1+\hat a\alpha}{1-\hat a\alpha}\,, \qquad
\cos\Theta=3\sqrt{3}\alpha\frac{\sqrt{1-\hat a^2\alpha^2}}{(1+\hat a\alpha)^2}\,,
\eeq
and we have assumed $\alpha>0$ for simplicity.
The condition $u_2=u_3$ gives the critical value of $\alpha$ corresponding to capture by the black hole.
For instance, the critical impact parameter for corotating orbits in the limit $\hat E\to\infty$ is given by (see, e.g., Ref. \cite{Frolov:1998wf})
\beq
\hat b_{\rm crit}=\hat a +8\cos^3\left(\frac13\arccos(-\hat a)\right)\,,
\eeq
so that $\alpha_{\rm crit}=1/\hat b_{\rm crit}$.

\subsection{Scattering angle}

\begin{figure*}
\[
\begin{array}{cc}
\includegraphics[scale=0.35]{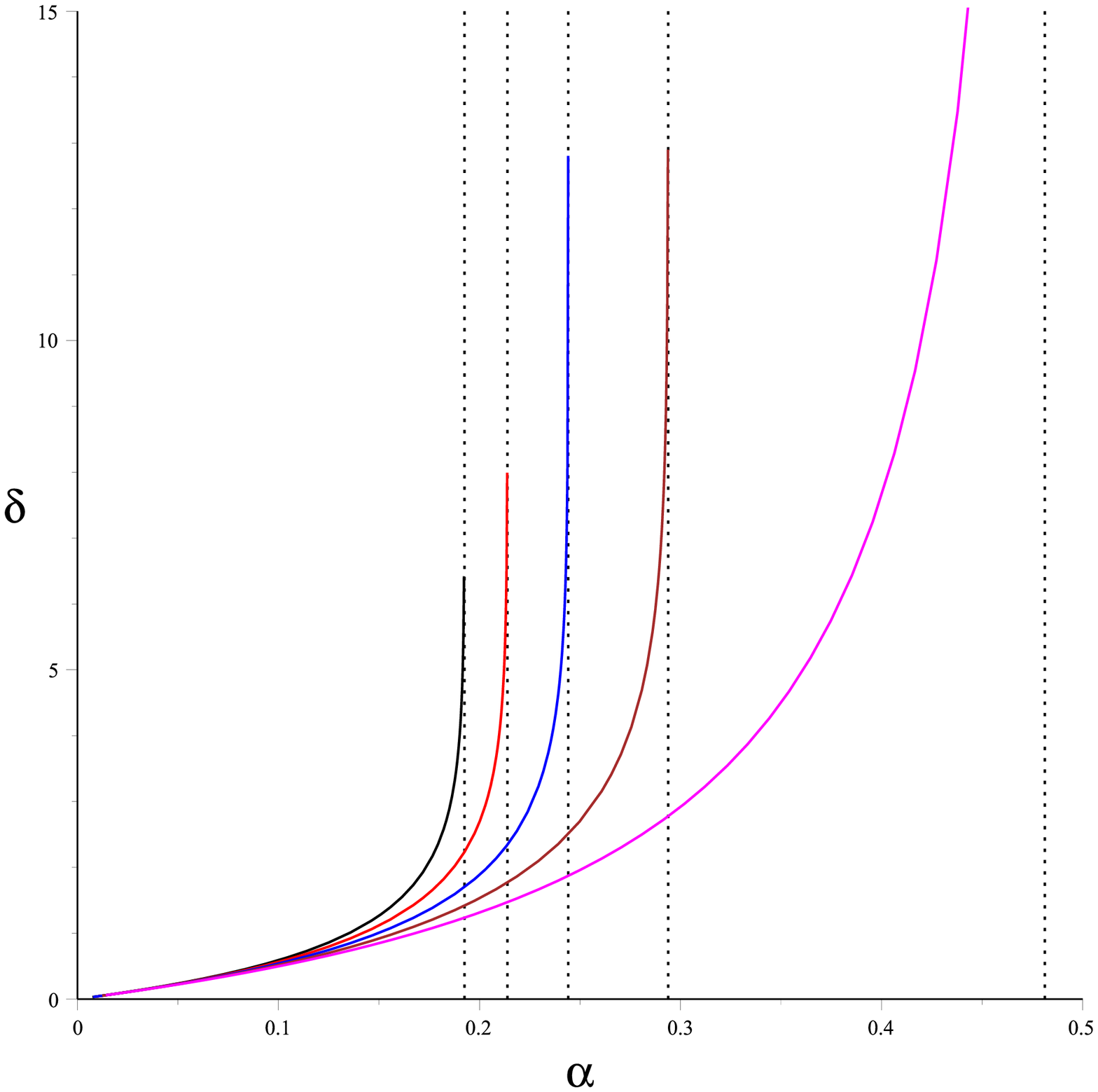}&\qquad 
\includegraphics[scale=0.35]{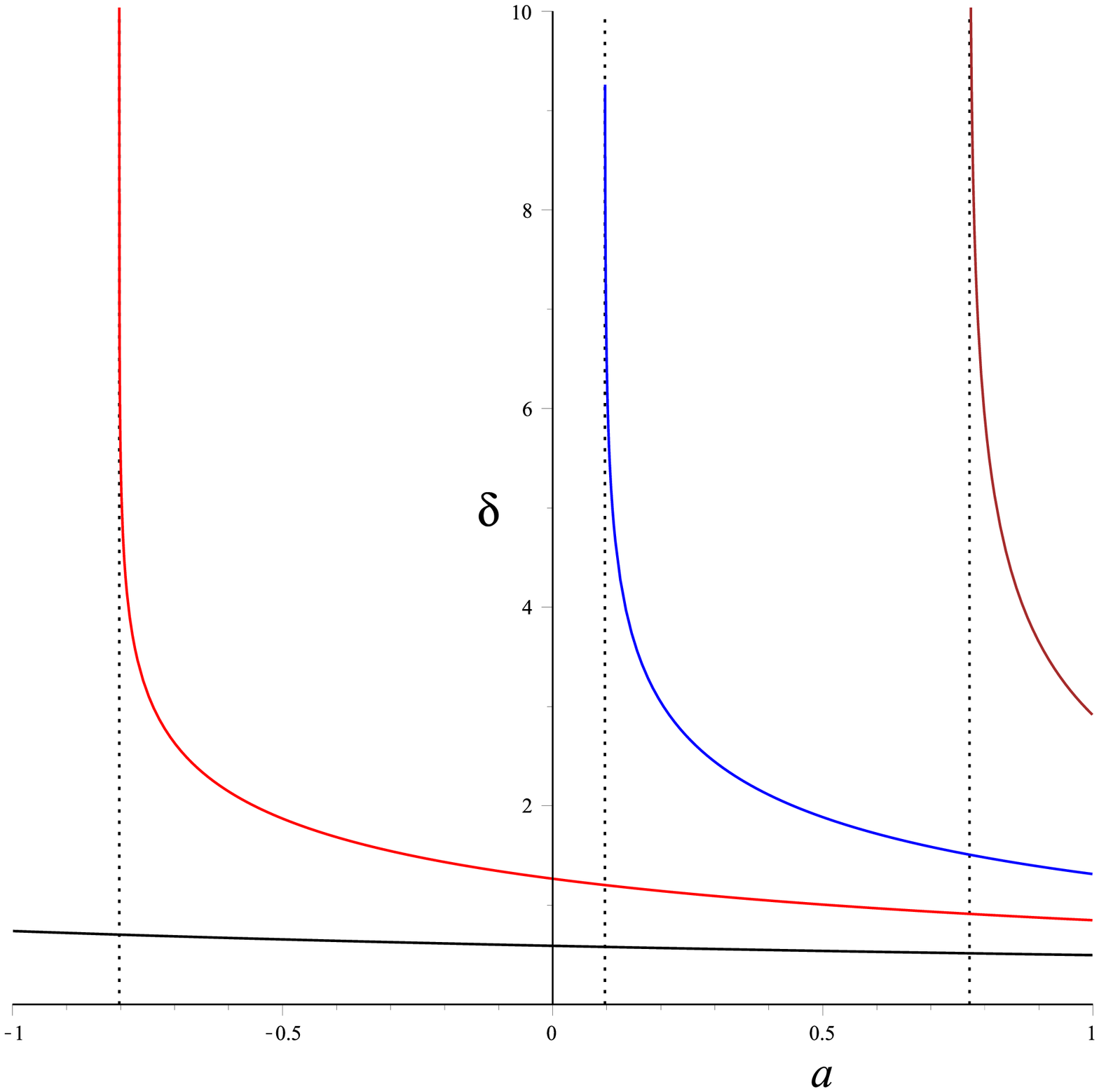}\cr
(a) & (b)
\end{array}
\]
\caption{The behavior of the deflection angle as a function of $\alpha$ in the Kerr spacetime is shown in panel (a) for selected values of $\hat a=[0,0.25,0.5,0.75,1]$.
The curves are ordered from left to right for increasing values of $\hat a$.
Dotted vertical lines show the corresponding critical values of $\alpha$ for capture by the black hole.
Panel (b) shows instead the behavior of the deflection angle as a function of $\hat a$ is shown for selected values of $\alpha=[0.1,0.15,0.2,0.3]$.
The curves are ordered from left to right for increasing values of $\alpha$.
Particles with $\alpha=0.1$ (lowest curve) are not captured by the black hole.
}
\label{fig:1}
\end{figure*}

The solution to Eq. \eqref{eqphidiu} can be expressed in terms of elliptic integrals as 

\beq
\label{phidiusol2}
\phi(u)=\frac{2\sqrt{2}}{\hat a^2(u_+-u_-)\sqrt{u_3-u_1}}
(A_+(u)-A_-(u))\,,
\eeq
where
\begin{eqnarray}
A_\pm(u)&=&\frac{u_4-u_\pm}{u_1-u_\pm}\left(\Pi(\psi,\beta_\pm,m)-\Pi(\beta_\pm,m)\right)
\,,\nonumber\\
\psi&=&\psi(u; u_1,u_2)=\sqrt{\frac{u-u_1}{u_2-u_1}}
\,,\nonumber\\
\beta_\pm&=&\frac{u_2-u_1}{u_\pm-u_1}=\frac{1}{[\psi(u_\pm; u_1,u_2)]^2}\,,\nonumber\\
m&=&\sqrt{\frac{u_2-u_1}{u_3-u_1}}=\frac{1}{\psi(u_3; u_1,u_2)}\,,
\end{eqnarray}
having assumed $\phi(u_2)=0$ at periastron.
Here 
\begin{eqnarray}
\Pi(\varphi,n,k)&=&\int_0^{\varphi}\frac{dz}{(1-n\sin^2z)\sqrt{1-k^2\sin^2z}}\,,\nonumber\\
\Pi(n,k)&=&\Pi(\pi/2,n,k)\,,
\end{eqnarray}
are the incomplete and complete elliptic integrals of the third kind, respectively~\cite{Gradshteyn}.
The total change in $\phi$ for the complete scattering process is then given by $2\phi(0)$ determined by Eq.~\eqref{phidiusol2} with $\psi=\psi(0)=\sqrt{-u_1/(u_2-u_1)}$, so that the deflection angle is $\delta(\hat a,\alpha)=2\phi(0)-\pi$.
Its behavior as a function of $\alpha$ is shown in Fig. \ref{fig:1} for selected values of $\hat a$.

For small values of $\alpha$, i.e., for large values of the impact parameter, we find

\begin{widetext}

\begin{eqnarray}
\label{deltakerr}
\delta(\hat a,\alpha)&=&4\alpha+\left(\frac{15}{4}\pi-4\hat a\right)\alpha^2
+\left(\frac{128}{3}-10\pi\hat a+4\hat a^2\right)\alpha^3
+\left(\frac{3465}{64}\pi-192\hat a+\frac{285}{16}\pi\hat a^2-4\hat a^3\right)\alpha^4\nonumber\\
&&
+\left(\frac{3584}{5}-\frac{693}{2}\pi\hat a+512\hat a^2-27\pi\hat a^3+4\hat a^4\right)\alpha^5\nonumber\\
&&
+\left(\frac{255255}{256}\pi-\frac{17920}{3}\hat a+\frac{79695}{64}\pi\hat a^2-\frac{3200}{3}\hat a^3+\frac{1195}{32}\pi\hat a^4-4\hat a^5\right)\alpha^6\nonumber\\
&&
+\left(\frac{98304}{7}-\frac{328185}{32}\pi\hat a+27136\hat a^2-\frac{13365}{4}\pi\hat a^3+1920\hat a^4-\frac{195}{4}\pi\hat a^5+4\hat a^6\right)\alpha^7\nonumber\\
&&
+\left(\frac{334639305}{16384}\pi-172032\hat a+\frac{115630515}{2048}\pi\hat a^2-89600\hat a^3+\frac{7661115}{1024}\pi\hat a^4-3136\hat a^5+\frac{15645}{256}\pi\hat a^6-4\hat a^7\right)\alpha^8\nonumber\\
&&
+O(\alpha^9)\,.
\end{eqnarray}

\end{widetext}
In the Schwarzschild limit we have $\beta_-=0$ and $\Pi(\psi,0,m)=F(\psi,m)$ and $\Pi(0,m)=K(m)$, so that 
\beq
\label{phiuschw}
\phi(u)=\frac{2\sqrt{2}}{\sqrt{u_3-u_1}}\left[ K(m)-F(\psi,m) \right]\,,
\eeq
where $F(\varphi,k)$ and $K(k)$ are the incomplete and complete elliptic integrals of the first kind, respectively, defined by
\beq
F(\varphi,k)=\int_0^{\varphi}\frac{dz}{\sqrt{1-k^2\sin^2z}}\,,\quad
K(k)=F(\pi/2,k)\,.
\eeq 
The roots $u_1$, $u_2$ and $u_3$ are given by Eqs. \eqref{roots} with $\bar u=1/6$ and $\Theta={\rm arccos}(3\sqrt{3}\alpha)$, whence
\begin{eqnarray}
u_2 -u_1 &=& \frac{1}{\sqrt{3}}\cos \left(\frac23{\rm arccos}(3\sqrt{3}\alpha) +\frac{\pi}{6}\right)\,,\nonumber\\
u_3-u_1 &=& \frac{1}{\sqrt{3}}\cos \left(\frac23{\rm arccos}(3\sqrt{3}\alpha) -\frac{\pi}{6}\right)\,.
\end{eqnarray}
The scattering angle \eqref{deltakerr} then becomes
\begin{eqnarray}
\delta(0,\alpha)&=&4\alpha+ \frac{15}{4}\pi \alpha^2
+ \frac{128}{3}  \alpha^3
+ \frac{3465}{64}\pi  \alpha^4\nonumber\\
&+&  \frac{3584}{5}  \alpha^5
+\frac{255255}{256}\pi  \alpha^6 
 +\frac{98304}{7} \alpha^7 \nonumber\\
&+& \frac{334639305}{16384}\pi  \alpha^8
+O(\alpha^9)\,,
\end{eqnarray}
and can be rewritten as 
\beq
\delta(0,\alpha)=\delta_{\not\pi}(0,\alpha) +\pi \delta_\pi(0,\alpha)\,,
\eeq
with
\begin{eqnarray}
\delta_{\not\pi}(0,\alpha)&=&4\alpha+  
 \frac{128}{3}  \alpha^3
+  \frac{3584}{5}  \alpha^5
 +\frac{98304}{7} \alpha^7 \nonumber\\
&&+O(\alpha^9)\,,\nonumber\\
\delta_\pi(0,\alpha)&=& \frac{15}{4} \alpha^2
+ \frac{3465}{64} \alpha^4 
+\frac{255255}{256} \alpha^6 \nonumber\\
&&+\frac{334639305}{16384} \alpha^8
+O(\alpha^9)\,.
\end{eqnarray}
In the Kerr case, instead, one recognizes that the $\pi$/non-$\pi$ structure is related to even/odd powers of $\hat a^i \alpha^j$, i.e., terms with even values of $i+j$ have a $\pi$.

                          
\subsection{Effects induced by the multipolar structure of the moving body}

The hyperbolike-like equatorial motion of a particle endowed with spin-induced quadrupolar structure has been investigated in Ref. \cite{Bini:2017pee} according to the Mathisson-Papapetrou-Dixon (MPD) model \cite{Mathisson:1937zz,Papapetrou:1951pa,tulc59,Dixon:1970zza,ehlers77}.  
In the case of aligned spin the signed magnitude $s$ of the spin vector is a constant of motion and the orbital equation reads
\begin{eqnarray}
\label{du_dphi_spin}
\left(\frac{du}{d\phi}\right)^2 &=& {\mathcal V}(u; \hat E,\hat L, \hat s)
+O(\hat s^3)\,,
\end{eqnarray}
with
\begin{widetext}
\begin{eqnarray} 
\label{du_dphi_spin2}
{\mathcal V}(u; \hat E,\hat L, \hat s)&= &\frac{\hat\Delta^2}{\hat w^2}\Bigg\{(\hat E^2-1)(1+\hat a^2u^2)
-\hat L^2u^2+(u+\hat x^2u^3)\Big(2+3\hat\Delta u^2\hat s^2\Big)
\nonumber\\&+&
\frac{\hat\Delta u^3}{\hat w}\left[-(\hat E\hat x+\hat a)\left[2\hat x\hat s+\left(2\hat E+\frac{3\hat a\hat x^2u^3}{\hat w}\right)\hat s^2\right]  +\hat x^3u^3\hat s^2\right.\nonumber\\&+&\left.
(C_Q-1)[(1-3\hat x^2u^2)\hat x(1-2u)+(1-9\hat x^2u^2)\hat E\hat a-6\hat x(\hat\Delta-\hat E^2)]\right]\Bigg\}\,,
\end{eqnarray}
\end{widetext}
where
\beq
\hat\Delta=1-2u+\hat a^2u^2\,,\quad 
\hat w=\hat L-2\hat xu\,.
\eeq
The dimensionless energy and angular momentum $\hat E$ and $\hat J$ as well as the test-body's orbital angular momentum $\hat L$ at infinity (and also $\hat x$) are now defined by using the conserved bare mass $m_0$, namely
\begin{eqnarray}
\label{ELs}
\hat E&=&\frac{E}{m_0},\qquad \hat L=\hat J-\hat E \hat s,\nonumber\\
\hat L&=&\frac{L}{m_0M},\qquad\hat s=\frac{s}{m_0M}\,.
\end{eqnarray}
Note that Eq. \eqref{du_dphi_spin2} has been generalized by including the polarizability parameter $C_Q$ related to the shape of the body (see, e.g., Refs. \cite{Laarakkers:1997hb,Steinhoff:2012rw,Hinderer:2013uwa,Bini:2015zya}) with respect to that of Ref. \cite{Bini:2017pee}, where only the simplest case $C_Q=1$ corresponding to a \lq\lq black-hole-like" extended body was considered. In particular in Ref. \cite{Laarakkers:1997hb} the (fitted) values of $C_Q$ for different equations of state of a rotating neutron star are explicitly  shown (see Table VII there).

In the ultrarelativistic limit and for small values of $\alpha$, defined in terms of $\hat J$,
\beq
\alpha\equiv \alpha_J=\frac{\sqrt{\hat E^2-1}}{\hat J}\,,
\eeq
the deflection angle turns out to be
\begin{eqnarray}
\label{defl_ang1}
\delta(\hat s,\hat a,\alpha) &=&\delta_{\rm K}(\hat a,\alpha)+(C_Q-1)\hat s^2\delta_{\hat s^2}(\hat a,\alpha)\nonumber\\
&&+\hat E^{-2}\delta^{(-2)}(\hat s,\hat a,\alpha)+O(\hat E^{-4})\,,
\end{eqnarray}
with 

\begin{widetext}

\begin{eqnarray} 
\label{defl_ang2}
\delta_{\hat s^2}(\hat a,\alpha)&=&4\alpha^3+\left(\frac{135}{16}\pi-12\hat a\right)\alpha^4
+\left(\frac{768}{5}-45\pi\hat a+24\hat a^2\right)\alpha^5
+\left(\frac{17325}{64}\pi-1152\hat a+\frac{4275}{32}\pi\hat a^2-40\hat a^3\right)\alpha^6\nonumber\\
&&
+\left(4608-\frac{10395}{4}\pi\hat a+4608\hat a^2-\frac{1215}{4}\pi\hat a^3+60\hat a^4\right)\alpha^7\nonumber\\
&&
+\left(\frac{16081065}{2048}\pi-53760\hat a+\frac{1673595}{128}\pi\hat a^2-13440\hat a^3+\frac{75285}{128}\pi\hat a^4-84\hat a^5\right)\alpha^8\nonumber\\
&&
+O(\alpha^9)\,.
\end{eqnarray}
Therefore, in the high-energy limit $\hat E\to \infty$ there is no contribution linear in spin to the scattering angle. Corrections to the spinless (geodesic) value start at the order $O(\hat s^2)$, provided that $C_Q\not=1$, i.e., for extended bodies which are not \lq\lq black-hole-like."
This circumstance reflects the peculiarity of the MPD model for these objects, as already pointed out in Ref. \cite{Bini:2015zya} for what concerns the alignment of the generalized momentum $P$ and the unit tangent vector $U$ to the world line representative of the body (see Eq. (4.21) there).

Corrections linear in spin appear at the order $\hat E^{-2}$. In fact, we find the following expression for $\delta^{(-2)}(\hat s,\hat a,\alpha)$
\beq
\delta^{(-2)}(\hat s,\hat a,\alpha)=\delta^{(-2)}_{\rm K}(\hat a,\alpha)+\hat s\delta^{(-2)}_{\hat s}(\hat a,\alpha)+(C_Q-1)\hat s^2\delta^{(-2)}_{\hat s^2}(\hat a,\alpha)\,,
\eeq
with
\begin{eqnarray} 
\delta^{(-2)}_{\rm K}(\hat a,\alpha)&=&
2\alpha+\left(3\pi-2\hat a\right)\alpha^2+\left(48-9\pi\hat a+2\hat a^2\right)\alpha^3
+\left(\frac{315}{4}\pi-240\hat a+\frac{33}{2}\pi\hat a^2-2\hat a^3\right)\alpha^4\nonumber\\
&&
+\left(1280-\frac{2205}{4}\pi\hat a+672\hat a^2-\frac{51}{2}\pi\hat a^3+2\hat a^4\right)\alpha^5\nonumber\\
&&
+\left(\frac{135135}{64}\pi-11520\hat a+\frac{16695}{8}\pi\hat a^2-1440\hat a^3+\frac{285}{8}\pi\hat a^4-2\hat a^5\right)\alpha^6\nonumber\\
&&
+\left(\frac{172032}{5}-\frac{1486485}{64}\pi\hat a+55040\hat a^2-\frac{46305}{8}\pi\hat a^3+2640\hat a^4-\frac{375}{8}\pi\hat a^5+2\hat a^6\right)\alpha^7\nonumber\\
&&
+\left(\frac{14549535}{256}\pi-\frac{2236416}{5}\hat a+\frac{17162145}{128}\pi\hat a^2-188160\hat a^3+\frac{424305}{32}\pi\hat a^4-4368\hat a^5+\frac{945}{16}\pi\hat a^6-2\hat a^7\right)\alpha^8\nonumber\\
&&
+O(\alpha^9)\,,\nonumber\\
\delta^{(-2)}_{\hat s}(\hat a,\alpha)&=&
2\alpha^2+\left(3\pi-4\hat a\right)\alpha^3+\left(48-\frac{27}{2}\pi\hat a+6\hat a^2\right)\alpha^4
+\left(\frac{315}{4}\pi-320\hat a+33\pi\hat a^2-8\hat a^3\right)\alpha^5\nonumber\\
&&
+\left(1280-\frac{11025}{16}\pi\hat a+1120\hat a^2-\frac{255}{4}\pi\hat a^3+10\hat a^4\right)\alpha^6\nonumber\\
&&
+\left(\frac{135135}{64}\pi-13824\hat a+\frac{50085}{16}\pi\hat a^2-2880\hat a^3+\frac{855}{8}\pi\hat a^4-12\hat a^5\right)\alpha^7\nonumber\\
&&
+\left(\frac{172032}{5}-\frac{3468465}{128}\pi\hat a+77056\hat a^2-\frac{324135}{32}\pi\hat a^3+6160\hat a^4-\frac{2625}{16}\pi\hat a^5+14\hat a^6\right)\alpha^8\nonumber\\
&&
+O(\alpha^9)\,,\nonumber\\
\delta^{(-2)}_{\hat s^2}(\hat a,\alpha)&=&
2\alpha^3+\left(9\pi-6\hat a\right)\alpha^4+\left(224-\frac{93}{2}\pi\hat a+12\hat a^2\right)\alpha^5
+\left(\frac{7875}{16}\pi-1696\hat a+\frac{1095}{8}\pi\hat a^2-20\hat a^3\right)\alpha^6\nonumber\\
&&
+\left(9984-\frac{38745}{8}\pi\hat a+6816\hat a^2-\frac{2475}{8}\pi\hat a^3+30\hat a^4\right)\alpha^7\nonumber\\
&&
+\left(\frac{315315}{16}\pi-120064\hat a+\frac{791595}{32}\pi\hat a^2-19936\hat a^3+\frac{9555}{16}\pi\hat a^4-42\hat a^5\right)\alpha^8\nonumber\\
&&
+O(\alpha^9)\,.
\end{eqnarray}
Again, terms quadratic in spin vanish for $C_Q=1$.

\end{widetext}

Let us recall that using the conserved energy and angular momentum  associated with the Killing vectors $\partial_t$ and $\partial_\phi$ of the Kerr spacetime evaluated at $r\to \infty$ one finds the relation (see Ref. \cite{Bini:2017pee}, footnote 1 on p. 8)
\beq
\label{JL}
\hat J =\hat L +\hat E \hat s\,.
\eeq
Let $b_L$ and $b_J$ be defined as
\beq
\hat b_L=\frac{\hat L}{\sqrt{\hat E^2-1}}\,,\qquad \hat b_J=\frac{\hat J}{\sqrt{\hat E^2-1}}\,.
\eeq
Inserting these expressions in Eq. \eqref{JL} leads to
\beq
\sqrt{\hat E^2-1} \hat b_J =\sqrt{\hat E^2-1} \hat b_L +\hat E \hat s\,,
\eeq
or
\beq
\hat b_J =\hat b_L +\frac{\hat E}{\sqrt{\hat E^2-1} } \hat s\,,
\eeq
which in the limit $\hat E \to \infty$ implies simply a shift by $\hat s$ in the two definitions, namely
\beq
\hat b_J =\hat b_L +  \hat s\,.
\eeq
Let us define then
\beq
\alpha_L =\frac{1}{\hat b_L}\,,\qquad \alpha_J =\frac{1}{\hat b_J}\,,
\eeq
so that
\beq
\alpha_L =\frac{\alpha_J}{1-\alpha_J \hat s}\approx \alpha_J+\alpha_J^2\hat s+\alpha_J^3\hat s^2+O(\hat s^3) \,.
\eeq
Consistently, Eq. \eqref{du_dphi_spin} implies $d\phi/du\to1/\alpha_L$ in the limit $u\to0$.
It is clear that our results \eqref{defl_ang1} and  \eqref{defl_ang2} could have been formulated by using $\alpha_L$ instead of $\alpha_J$. 
In terms of $\alpha_L$ the spin corrections to the scattering angle would then appear already at leading order, and not at order $O(\hat E^{-2})$. In fact, in this case we find
\begin{eqnarray}
\delta(\hat s,\hat a,\alpha_L) &=&\delta_{\rm K}(\hat a,\alpha_L)
+\hat s\delta_{\hat s}(\hat a,\alpha_L)\nonumber\\
&&
+\hat s^2\delta_{\hat s^2}(\hat a,\alpha_L)
+O(\hat E^{-2})\,,
\end{eqnarray}
with
\begin{eqnarray}
\delta_{\hat s}(\hat a,\alpha_L)&=&
-4\alpha_L^2+\left(8\hat a-\frac{15}{2}\pi\right)\alpha_L^3\nonumber\\
&&
+(-128+30\pi\hat a-12\hat a^2)\alpha_L^4\nonumber\\
&&
+\left(-\frac{3465}{16}\pi+768\hat a-\frac{285}{4}\pi\hat a^2+16\hat a^3\right)\alpha_L^5\nonumber\\
&&
+O(\alpha_L^6)
\,,\nonumber\\
\delta_{\hat s^2}(\hat a,\alpha_L)&=&
4C_Q\alpha_L^3
+\left[\frac{45}{4}\pi-12\hat a\right.\nonumber\\
&&\left.
+\left(\frac{135}{16}\pi-12\hat a\right)(C_Q-1)\right]\alpha_L^4\nonumber\\
&&
+\left[256-60\pi\hat a+24\hat a^2\right.\nonumber\\
&&\left.
+\left(\frac{768}{5}-45\pi\hat a+24\hat a^2\right)(C_Q-1)\right]\alpha_L^5\nonumber\\
&&
+O(\alpha_L^6)
\,,
\end{eqnarray}
which agree with Eq. (5.23) of Ref. \cite{Barrabes:2003ey} (see also Ref. \cite{Barrabes:2004gn}), where a completely different method were actually used.

\subsection{Two-body scattering angle in PM theory}

In a recent work Damour has shown how to translate the ultra high-energy quantum scattering results of Amati, Ciafaloni and Veneziano \cite{Amati:1987wq,Amati:1987uf} into a classical relativistic gravitational two-body scattering angle in PM perturbation theory \cite{Damour:2017zjx}.
The scattering process between the two (nonrotating) bodies is equivalently described in terms of the scattering of a massless particle in a modified Schwarzschild metric of the form  
\beq
\label{modschw}
ds^2 =-f_t dt^2 +f_r^{-1}dr^2 +r^2 (d\theta^2+\sin^2\theta d\phi^2)\,,
\eeq
with
\begin{eqnarray}
f_t(u) &=& \left(1-2u\right)\left(1+\frac{15}{2}u^2 -18 u^3+ \frac{1845}{16}u^4+\ldots \right)\,, \nonumber\\
f_r(u) &=& (1-2u) \,,
\end{eqnarray}
where $u=M/r$ is a dimensionless \lq\lq inverse radial" variable.
The scattering equation in this case reduces to
\beq
\label{dphidu2pm}
\frac{d\phi}{du}=\pm\sqrt{\frac{N(u)}{D(u)}} \,,
\eeq
where
\begin{eqnarray}
N(u)&=& 1+\frac{15}{2}u^2-18u^3+\frac{1845}{16}u^4\,,\nonumber\\
D(u)&=&\alpha^2-u^2\left(1-2u+\frac{15}{2}u^2-33u^3\right.\nonumber\\
&&\left.
+\frac{2421}{16}u^4-\frac{1845}{8}u^5\right)
\,,
\end{eqnarray} 
and the scattering angle results in
\beq
\label{deltadam}
\delta(\alpha) =  4\alpha+\frac{56}{3}\alpha^3+O(\alpha^5)\,,
\eeq
with vanishing  both the coefficients at $O(\alpha^2)$ and $O(\alpha^4)$, as shown in Ref. \cite{Damour:2017zjx}.
Note that to integrate Eq. \eqref{dphidu2pm} we have used the prescription given there (see Eqs. (7.23)-(7.24)).

A general modified Schwarzschild metric has the form \eqref{modschw}, with functions
\begin{eqnarray}
f_t(u) &=& (1-2u)\left(1+c_t^{(2)} u^3 +c_t^{(3)} u^3+c_t^{(4)} u^4+\ldots\right)\,, \nonumber\\
f_r(u) &=& (1-2u)\left(1+c_r^{(2)} u^2+c_r^{(3)} u^3+c_r^{(4)} u^4+\ldots\right) \,.\nonumber\\
\end{eqnarray}
This includes spherically symmetric solutions from extended theories of gravity (see e.g., Refs. \cite{Maeda:2006hj,Sotiriou:2013qea,Mukherjee:2017fqz}), the functions $f_t(u)$ and $f_r(u)$ being parametrized by a number of coefficients generically associated with scalar charges.
The scattering angle in the high energy limit turns out to be
\begin{eqnarray}
\delta(c_t^{(i)},c_r^{(i)},\alpha)&=&4\alpha
+\left(\frac{15}{4}-\frac14c_r^{(2)}-\frac12c_t^{(2)}\right)\pi\alpha^2\nonumber\\
&+&
\left(\frac{128}{3}-8c_t^{(2)}-\frac83c_r^{(2)}-\frac23c_r^{(3)}\right. \nonumber\\
&-&\left. 2c_t^{(3)}\right)\alpha^3+\left(\frac{3465}{64}+\frac98(c_t^{(2)})^2\right.\nonumber\\
&+&
\left. \frac{9}{64}(c_r^{(2)})^2-\frac{105}{8}c_t^{(2)}-\frac{105}{32}c_r^{(2)}\right.\nonumber\\
&+&
\frac38c_r^{(2)}c_t^{(2)}-\frac{15}{16}c_r^{(3)}-\frac{15}{4}c_t^{(3)}
\nonumber\\
&-&\left.
\frac{3}{16}c_r^{(4)}-\frac{3}{4}c_t^{(4)}\right)\pi\alpha^4
+O(\alpha^5)\,.
\end{eqnarray}
Note that the absence in $\delta$ of even powers of $\alpha$ is compatible with the choice
\begin{eqnarray}
\label{ct24sol}
c_t^{(2)} &=&  \frac{15}{2}-\frac12 c_r^{(2)}\,, \nonumber\\
c_t^{(4)}&=&  \frac{405}{16}-5c_t^{(3)}-\frac{25}{8}c_r^{(2)}+\frac{5}{16}(c_r^{(2)})^2\nonumber\\
&&-\frac{5}{4}c_r^{(3)}-\frac14 c_r^{(4)}\,.
\end{eqnarray}
The result \eqref{deltadam} is reproduced by setting $c_r^{(2)}=c_r^{(3)}=c_r^{(4)}=0$ and 
$c_t^{(2)}=\frac{15}{2}$, $c_t^{(3)}= -18$, $c_t^{(4)}=\frac{1845}{16}$.
In fact, for $c_r^{(2)}=c_r^{(3)}=c_r^{(4)}=0$ the coefficients \eqref{ct24sol} become
\beq
c_t^{(2)} =  \frac{15}{2}\,, \qquad
c_t^{(4)}= \frac{405}{16}-5c_t^{(3)} \,,
\eeq
which agree with Eq. \eqref{deltadam} with the choice $c_t^{(3)}=-18$, implying in turn $c_t^{(4)}=\frac{1845}{16}$.

\section{Hyperbolic-like equatorial motion in the Hartle-Thorne spacetime}

The Hartle-Thorne metric describing the exterior field of a slowly rotating/slightly deformed object is given by~\cite{Hartle:1968si}

\begin{figure}
\centering
\includegraphics[scale=0.4]{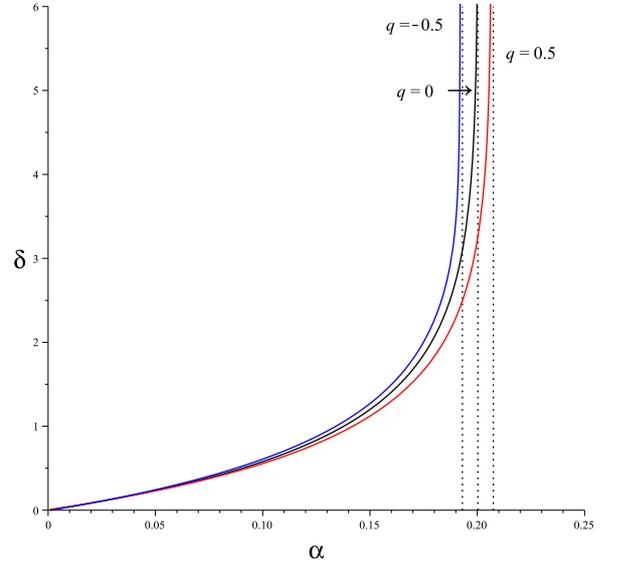}
\caption{The behavior of the deflection angle as a function of $\alpha$ in the HT spacetime is shown for $\hat a=0.1$ in both prolate ($q=0.5$) and oblate ($q=-0.5$) cases.
Dotted vertical lines show the corresponding critical values of $\alpha$ for capture by the star.
The values of the quadrupole parameter have been exaggerated to better show the effect. 
}
\label{fig:2}
\end{figure}

\begin{widetext}

\begin{eqnarray}
\label{HTmet}
ds^2&=&-\left(1-\frac{2{\mathcal M}}{R}\right)\left[
1+2k_1P_2(\cos\Theta)+2\left(1-\frac{2{\mathcal M}}{R}\right)^{-1}\frac{{\mathcal J}^2}{R^4}(2\cos^2\Theta-1)\right]dt^2\nonumber\\
&&+\left(1-\frac{2{\mathcal M}}{R}\right)^{-1}\left[
1-2k_2P_2(\cos\Theta)-2\left(1-\frac{2{\mathcal M}}{R}\right)^{-1}\frac{{\mathcal J}^2}{R^4}\right]dR^2\nonumber\\
&&+R^2(d\Theta^2+\sin^2\Theta d\phi^2)[1-2k_3P_2(\cos\Theta)]-4\frac{{\mathcal J}}{R}\sin^2\Theta dt d\phi\ ,
\end{eqnarray}
where 
\begin{eqnarray}
k_1&=&\frac{{\mathcal J}^2}{{\mathcal M}R^3}\left(1+\frac{{\mathcal M}}{R}\right)-\frac58\frac{{\mathcal Q}-{\mathcal J}^2/{\mathcal M}}{{\mathcal M}^3}Q_2^2\left(\frac{R}{{\mathcal M}}-1\right)\ , \nonumber\\
k_2&=&k_1-\frac{6{\mathcal J}^2}{R^4}\ , \nonumber\\
k_3&=&k_1+\frac{{\mathcal J}^2}{R^4}-\frac54\frac{{\mathcal Q}-{\mathcal J}^2/{\mathcal M}}{{\mathcal M}^2R}\left(1-\frac{2{\mathcal M}}{R}\right)^{-1/2}Q_2^1\left(\frac{R}{\mathcal M}-1\right)\ . 
\end{eqnarray}
Here $Q_l^m$ are the associated Legendre functions of the second kind and the constants ${\mathcal M}$, ${\mathcal J}$ and ${\mathcal Q}$ are the total mass, angular momentum and mass quadrupole moment of the rotating source respectively.

Let us introduce Boyer-Lindquist coordinates $(t,r,\theta,\phi)$ through the transformation \cite{Hartle:1968si,Glampedakis:2005cf,Bini:2009cg}
\beq
t=t\ , \qquad
R=R(r,\theta)\ , \qquad 
\Theta=\Theta(r,\theta)\ , \qquad
\phi=\phi\ 
\eeq
with 
\begin{eqnarray}
R&=&r\left\{1+\frac12\frac{{\mathcal J}^2}{{\mathcal M}^2r^2}\left[\left(1+\frac{2\mathcal M}{r}\right)\left(1-\frac{\mathcal M}{r}\right)-\cos^2\theta\left(1-\frac{2\mathcal M}{r}\right)\left(1+\frac{3\mathcal M}{r}\right)\right]\right\}
\,,\nonumber\\
\Theta&=&\theta+\sin\theta\cos\theta\frac12\frac{{\mathcal J}^2}{{\mathcal M}^2r^2}\left(1+\frac{2\mathcal M}{r}\right)\,.
\end{eqnarray}
When $\mathcal Q=\mathcal J^2/M$ and $\mathcal J=Ma$ (with $\mathcal M=M$) the metric \eqref{HTmet} then reduces to the Kerr metric \eqref{kerrmet} up to second order in the rotation parameter $\hat a=a/M$.
Therefore it is convenient to introduce the dimensionless quadrupole parameter $q$ by $\mathcal Q=\mathcal J^2/M+qM^3=(\hat a^2+q)M^3$, representing the deviation from the (slowly rotating) Kerr solution due to the mass quadupole moment of the source.
Negative values of $q$ correspond to oblate configurations, whereas positive values to prolate ones, relative to the $z$-axis.
The HT metric \eqref{HTmet} written in BL coordinates thus becomes
\beq
g^{\rm HT}_{\alpha\beta}=g^{{\rm K},\hat a^2}_{\alpha\beta}+qg^{q}_{\alpha\beta}+O(\hat a^3,q^2)\,,
\eeq
and is accurate to second order in $\hat a$ and to first order in $q$ (terms like $\hat aq$ and higher powers have also been neglected).
The nonvanishing components of the quadrupolar correction are given by
\begin{eqnarray}
g^{q}_{tt}&=&\left(1-\frac{2M}{r}\right)^2g^{q}_{rr}
=\frac54P_2(\cos\theta)\left(1-\frac{2M}{r}\right)Q_2^2\left(\frac{r}{M}-1\right)
\,,\nonumber\\
g^{q}_{\theta\theta}&=&\frac{g^{q}_{\phi\phi}}{\sin^2\theta}
=\frac54P_2(\cos\theta)r^2\left[\frac{2M}{r}\left(1-\frac{2M}{r}\right)^{-1/2}Q_2^1\left(\frac{r}{M}-1\right)+Q_2^2\left(\frac{r}{M}-1\right)\right]
\,.
\end{eqnarray}

The geodesic equations in the equatorial plane are separable \cite{Glampedakis:2005cf}, and can be written as
\begin{eqnarray}
\label{geosHT}
\frac{dt}{d\tau} &=&\left(\frac{dt}{d\tau}\right)_{{\rm K},\hat a^2}
-\frac58q\hat E\left(1-\frac{2M}{r}\right)^{-1}Q_2^2\left(\frac{r}{M}-1\right) 
\,, \nonumber\\
\left(\frac{dr}{d\tau}\right)^2  &=&\left(\frac{dr}{d\tau}\right)^2_{{\rm K},\hat a^2}
-\frac58q\left(1-\frac{2M}{r}\right)\left[\left(1+2\frac{\hat L^2}{r^2}\right)Q_2^2\left(\frac{r}{M}-1\right)+\frac{2M\hat L^2}{r^3}\left(1-\frac{2M}{r}\right)^{-1/2}Q_2^1\left(\frac{r}{M}-1\right)\right]
\,, \nonumber\\
\frac{d\phi}{d\tau}&=&\left(\frac{d\phi}{d\tau}\right)_{{\rm K},\hat a^2} 
+\frac58q\frac{\hat L}{r^2}\left[Q_2^2\left(\frac{r}{M}-1\right)+\frac{2M}{r}\left(1-\frac{2M}{r}\right)^{-1/2}Q_2^1\left(\frac{r}{M}-1\right)\right] 
\,,
\end{eqnarray} 
where $({dx^\alpha}/{d\tau})_{{\rm K},\hat a^2}$ are given by the Kerr values \eqref{eqsmotionequatorial}, truncated at the second order in $\hat a$.

The orbital equation is thus given by
\begin{eqnarray} 
\label{dudphi_HT}
\left(\frac{du}{d\phi}\right)^2&=&\left(\frac{du}{d\phi}\right)^2_{{\rm K},\hat a^2}
-\frac58q\left[\frac{2u}{\sqrt{1-2u}}\left(2u^3-u^2+2\frac{M^2}{\hat L^2}(2u+\hat E^2-1)\right)Q_2^1\left(\frac{1-u}{u}\right)\right.\nonumber\\
&&\left.
+\frac{M^2}{\hat L^2}(2u+2\hat E^2-1)Q_2^2\left(\frac{1-u}{u}\right)\right]
\,,
\end{eqnarray}

\end{widetext}
and in the ultrarelativistic limit reduces to
\beq
\label{dphi_du_HT}
\frac{d \phi}{d u}=\left(\frac{d\phi}{du}\right)_{{\rm K},\hat a^2}
+q\left(\frac{d\phi}{du}\right)_{q}
+O(\hat E^{-2})\,,
\eeq
with 
\begin{eqnarray} 
\left(\frac{d\phi}{du}\right)_{q}&=&
\frac{15}{16}\frac{2u^5-3u^4+u^3+2u^2\alpha^2-\alpha^2}{u^2(2u^3-u^2+\alpha^2)^{3/2}}\ln(1-2u)\nonumber\\
&&
+\frac{5}{8}\frac{u^5-6u^4+3u^3+2u^2\alpha^2-3u\alpha^2-3\alpha^2}{u(2u^3-u^2+\alpha^2)^{3/2}}
\,.\nonumber\\
\end{eqnarray}
Turning points for radial motion are the roots of the rhs of Eq. \eqref{dudphi_HT}, which can be written as first order corrections in $q$ to the Kerr ones (see Eq. \eqref{roots}), i.e., $u_i=u_i^{{\rm K},\hat a^2}+qu_i^{q}$, with
\begin{eqnarray} 
u_i^{q}&=&\frac{5}{16}\frac{3\alpha^2+(u_i^{\rm S})^2+12u_i^{\rm S}-6}{3u_i^{\rm S}-1}\nonumber\\
&&
+\frac{15}{16}\alpha^2\frac{(u_i^{\rm S})^2+u_i^{\rm S}-1}{(u_i^{\rm S})^3(3u_i^{\rm S}-1)}\ln(1-2u_i^{\rm S})\,,
\end{eqnarray}
$u_i^{\rm S}$ denoting the corresponding roots for the Schwarzschild case ($\hat a=0$).
The condition for capture is thus modified as $\alpha_{\rm crit}=\alpha_{\rm crit}^{{\rm K},\hat a^2}+q\alpha_{\rm crit}^q$, with
\beq
\alpha_{\rm crit}^q=\frac{5}{18}\sqrt{3}\left(\frac{15}{16}\ln(3)-1\right)\,.
\eeq
Equation \eqref{dphi_du_HT} can be numerically integrated to yield the deflection angle as a function of $(q, \hat a, \alpha)$.
Its behavior as a function of $\alpha$ for a fixed value of $\hat a$ is shown in Fig. \ref{fig:2} in both prolate ($q>0$) and oblate ($q<0$) cases.
For small values of $\alpha$ the deflection angle turns out to be
\beq
\delta(q,\hat a,\alpha)=\delta_{{\rm K}(\hat a,\alpha),\hat a^2}+q\delta_q(\alpha)\,,
\eeq
with 
\begin{eqnarray} 
\delta_q(\alpha)&=&-4\alpha^3-\frac{375}{32}\pi\alpha^4-\frac{1760}{7}\alpha^5-\frac{62625}{128}\pi\alpha^6\nonumber\\
&&
-\frac{186880}{21}\alpha^7-\frac{65131185}{4096}\pi\alpha^8
+O(\alpha^9)\,.
\end{eqnarray}

                          
\section{Discussion}

We have studied the high-energy scattering of an extended body endowed with structure up to the quadrupole by a Kerr black hole as well as of a pointlike particle in the field of a (slowly) rotating (slightly) deformed source endowed with a mass quadrupole moment, as described by the (approximate) Hartle-Thorne solution.
We have analytically computed the (conservative) scattering angle as a power series expansion in the (dimensionless) inverse impact parameter $\alpha$, kept as a constant in this regime.
It turns out to depend on the parameters characterizing the source (dimensionless intrinsic angular momentum $\hat a$ and quadrupole parameter $q$) as well as the scattered body (dimensionless spin parameter $\hat s$ and polarizability constant $C_Q$). It can be formally written as
\beq
\delta=\delta_{\rm geo}+\delta_{\rm struct}\,,
\eeq
where $\delta_{\rm geo}(\hat a, q, \alpha)$ contains information on the background source only, the scattered particle moving along a hyperbolic-like geodesic orbit, and 
\beq
\delta_{\rm struct}(\hat a, \hat s, C_Q, \alpha)=\delta_{\rm spin}(\hat s; \hat a,\alpha)+\delta_{{\rm spin}^2}(\hat s^2;\hat a, C_Q,\alpha)\,,
\eeq
depending also on the constitutive parameters of the scattered extended body moving along a non-geodesic path with spin aligned with that of the background source.

The dynamics of structured particles has been described according to the MPD model, as it is customary.
Their quadrupole moment is taken to be proportional to the square of the spin by a constant parameter $C_Q$, with $C_Q=1$ when the particle is \lq\lq black hole-like." Other interesting cases with $C_Q\not =1$ arise when the scattered particle is instead a neutron star, the value of $C_Q$ depending on the equation of state.  
A direct measurement of $\delta_{\rm struct}$ can then be used to constrain the equation of state of the  extended body undergoing the scattering process, if its spin is known. This information will also play a role in other contexts, for example when one is reconstructing a gravitational wave signal from a binary system with one component being  a neutron star with a certain equation of state.

In principle, these  studies are preliminar to a fully perturbative analysis where also the backreaction effects of the particle on the background is taken into account. This remains an open issue for future works.

\section*{Acknowledgments}
The authors thank T. Damour for useful discussions.
D.B. thanks the Naples Section of the Italian Istituto Nazionale di Fisica Nucleare (INFN) and the International Center for Relativistic Astrophysics Network (ICRANet) for partial support.

\end{document}